\documentclass[aps,prapplied,reprint,amsmath,amssymb,floatfix,longbibliography,raggedbottom]{revtex4-2}

\usepackage{graphicx}
\usepackage{bm}
\usepackage{booktabs}
\usepackage{placeins}
\usepackage{xcolor}
\usepackage{hyperref}
\hypersetup{colorlinks=true,citecolor=blue,urlcolor=blue,linkcolor=blue}

\makeatletter
\newcommand{\widefigurecaption}[1]{%
  \refstepcounter{figure}%
  \@makecaption{\fnum@figure}{#1}%
}
\newenvironment{plainwidetext}{%
  \par\ignorespaces
  \onecolumngrid
  \vskip6\p@
}{%
  \par
  \vskip6\p@
  \twocolumngrid\global\@ignoretrue
  \@endpetrue
}
\makeatother

\makeatletter
\def\ps@footerpage{%
  \let\@oddhead\@empty
  \let\@evenhead\@empty
  \def\@oddfoot{\hfil\thepage\hfil}%
  \let\@evenfoot\@oddfoot
}
\makeatother

\begin{document}
\let\savedapsmaketitle\maketitle
\let\savedapsauthor\author
\let\savedapsemail\email
\let\savedapsaffiliation\affiliation
\pagestyle{footerpage}

\title{Multiscale Analysis of Electrically Tunable Reflection Zeros in Intersubband Polaritonic Metasurfaces}

\author{Inyong Hwang}
\email{makki5068@gmail.com}
\affiliation{Independent Researcher, Hwaseong-si, Gyeonggi-do, Republic of Korea}

\date{August 17, 2026}

\begin{abstract}
Electrical tuning of intersubband-polaritonic metasurfaces cannot be fully
identified from reflectance alone because intensity obscures the complex
pole--zero dynamics governing critical coupling. We establish a
field-consistent multiscale analysis that connects self-consistent
Schr\"odinger--Poisson quantum states, complex rigorous coupled-wave spectra,
and constrained one-port temporal coupled-mode theory across 21 electric
fields. A single zero-field complex calibration fixes the shared photonic and
phase-delay baseline, after which the effective coupling is the only
field-dependent fitted parameter. The model reproduces the full-wave spectra
with a complex-amplitude RMSE of 0.0255--0.0288 and is independently
cross-checked through microscopic residue and overlap calculations. Complex
poles describe internal polariton hybridization, whereas reflection zeros
expose the external radiative--dissipative balance. The quantum-confined Stark
effect drives the lower- and upper-polariton zeros through the real-energy
axis in opposite directions, yielding two branch-resolved critical-coupling
transitions that agree with direct full-wave cross-checks.
\end{abstract}

\maketitle
\thispagestyle{footerpage}

\section{Introduction}

Intersubband transitions (ISBTs) are optical transitions between electron
subbands confined along the semiconductor growth direction. Their energies,
wave functions, and dipole moments are electrically tunable, while the
selection rule requires an electric-field component normal to the
quantum-well layers \cite{ando1982,helm2000}. Metasurface resonators provide
this out-of-plane field and can hybridize with an ISBT to form lower- and
upper-polariton modes. Strong and ultrastrong intersubband coupling has been
observed in microcavities and deeply subwavelength resonators
\cite{dini2003,anappara2009,todorov2010,geiser2012,benz2013}.

Electrical control has progressed from tunable polaritonic reflection to
phase modulation, nonlinear switching, and beam manipulation
\cite{lee2014,chung2023,mann2021,yu2022,cotrufo2024}. In particular,
electrically phase-tunable InGaAs/InAlAs MQW plasmonic metasurfaces have been
fabricated and measured \cite{chung2023}. These experiments establish the
device feasibility of the platform considered here. They do not, however,
identify the complex reflection zeros that organize critical coupling. The
quantum-confined Stark effect (QCSE) also changes the transition energy,
dipole matrix element, oscillator strength, population difference, and
dephasing, so a rigidly shifted Lorentz oscillator is insufficient for a
field-consistent reduction.

An internal two-mode Hamiltonian and an external reflection coefficient are
distinct objects. Complex poles describe the internal polaritons, whereas
reflection also depends on the direct path, radiative leakage, dissipation,
and scattering zeros. Critical coupling is a real-frequency reflection zero
and cannot be inferred from spectral splitting alone. This distinction
underlies coherent perfect absorption and polaritonic critical coupling
\cite{zanotto2014,zanotto2016,chong2010,zhang2022,sweeney2020,krasnok2019}.
Temporal coupled-mode theory (TCMT) provides the corresponding input--output
description \cite{fan2003,suh2004,ruan2010,maksimov2025}, but fieldwise fits are useful only
when their shared and field-dependent parameters are identifiable.

Here we connect a self-consistent microscopic QCSE trajectory to the
branch-resolved pole--zero evolution of a one-port intersubband-polaritonic
metasurface. Schr\"odinger--Poisson (S--P) calculations supply the quantum
inputs, complex rigorous coupled-wave analysis (RCWA) supplies the reference
scattering response, and a constrained two-oscillator TCMT shares one
photonic and phase-delay baseline over 21 electric fields. After one
zero-field complex calibration, only the effective coupling
\(g_{\mathrm{fit}}(F)\) is fitted at nonzero field. The reduction separates
internal poles from external zeros and reveals oppositely directed real-axis
crossings of the lower- and upper-polariton zeros. Direct complex-RCWA,
bare-cavity, microscopic-overlap, multistart, and fitting-window checks test
the result without adding field-dependent fit parameters.

\section{Physical system and multiscale modeling}

\begin{figure}[!t]
\centering
\includegraphics[width=\columnwidth]{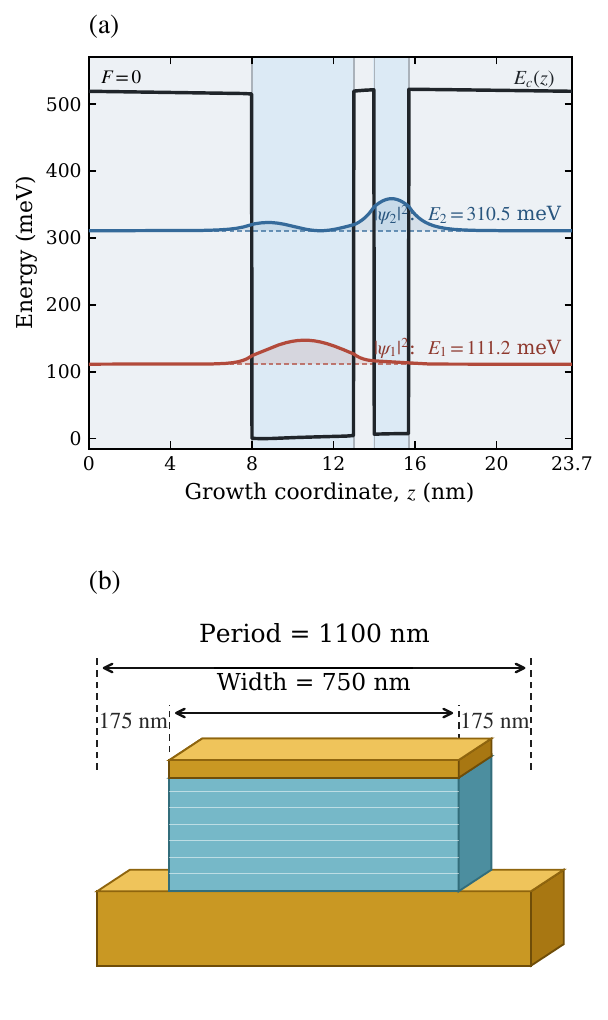}
\caption{(a) Zero-field self-consistent conduction-band edge and probability
densities of the two lowest states in the asymmetric InGaAs/InAlAs double
quantum well. (b) Spacer-free one-dimensional metal--insulator--metal grating
unit cell used for RCWA, with a co-patterned top-Au resonator and MQW active
region above a continuous Au ground plane.}
\label{fig:workflow}
\end{figure}

Fig.~\ref{fig:workflow} connects the quantum and electromagnetic scales. The
\(23.7~\mathrm{nm}\) asymmetric double-well unit contains, in sequence,
\(8.0~\mathrm{nm}\) InAlAs, \(5.0~\mathrm{nm}\) doped InGaAs,
\(1.0~\mathrm{nm}\) InAlAs, \(1.7~\mathrm{nm}\) InGaAs, and
\(8.0~\mathrm{nm}\) InAlAs. Its donor sheet density is
\(N_s=2.5\times10^{12}~\mathrm{cm^{-2}}\), and the optical transition is
between the two lowest overlap-tracked states. The growth coordinate \(z\)
is also the ISBT dipole direction.

\begin{samepage}
Fig.~\ref{fig:workflow}(a) shows that asymmetric confinement spatially
distinguishes the two states while preserving finite intersubband overlap;
bias-driven redistribution generates the QCSE trajectory discussed below.
Fig.~\ref{fig:workflow}(b) defines the optical unit cell.
\end{samepage}

The S--P calculation resolves one exact microscopic period. RCWA represents
eight periods by a nominal \(200~\mathrm{nm}\) homogenized MQW layer; the
unrounded microscopic thickness is \(189.6~\mathrm{nm}\). The distinction is
intentional: the former is the fabricated optical-layer thickness used in
the device model, whereas the latter is the sum of eight ideal quantum units.
It prevents a rounded device dimension from being mistaken for a microscopic
well parameter. The optical cell is
a spacer-free metal--insulator--metal grating with a \(50~\mathrm{nm}\)
top-Au resonator, a \(200~\mathrm{nm}\) Au ground plane, a
\(750~\mathrm{nm}\) patterned width, and a \(1100~\mathrm{nm}\) period. This
MQW-loaded, metal-backed resonator follows an experimentally established
intersubband-polaritonic device concept \cite{leeNature2014,chung2023}; related
full-wave and reduced descriptions of metasurface--ISBT coupling are given in
Refs.~\cite{gabbay2012,campione2014}. The ground plane makes the structure
effectively one port: over the full grid, transmission remains below
\(6.68\times10^{-9}\) and the energy-balance residual below \(3\times10^{-16}\).
The RCWA calculation retains the complex specular coefficient \(S_{11}\)
using a stable Fourier-modal formulation \cite{moharam1995}.
The incident polarization is along the grating-coupled in-plane direction;
the resonator converts it into the strong local \(E_z\) required by the ISBT
selection rule. Spectra are sampled uniformly in wavelength, but all pole,
zero, linewidth, and coupling quantities are evaluated in photon energy so
that every term in the reduced equations has the same unit.

\begin{samepage}
The field-dependent quantum inputs follow from the self-consistent S--P
eigenproblem. Within a one-dimensional effective-mass description, the
conduction-band electron envelope wavefunctions satisfy the eigenvalue
problem
\begin{equation}
\begin{split}
\biggl[
&-\frac{\hbar^2}{2}\frac{d}{dz}
\frac{1}{m^\ast(z,E_n)}\frac{d}{dz}
\\
&+V_c(z)+eF(z-z_c)-e\phi(z)
\biggr]\psi_n(z)
=E_n\psi_n(z)
\end{split}
\label{eq:sch}
\end{equation}
\end{samepage}
Here \(m^\ast(z,E_n)\) is the layer- and energy-dependent mass in the
BenDaniel--Duke operator \cite{bendaniel1966}; \(V_c\), \(F\), and \(\phi\)
are the unperturbed band edge, signed physical field, and Hartree potential.
The convention is \(U_{\mathrm{ext}}=eF(z-z_c)\), with \(e>0\) and positive
\(F\) along \(+\hat{\mathbf z}\). The Hartree potential is updated
self-consistently from the occupied-electron and ionized-donor densities;
\(\psi_n\) and \(m^{\ast-1}\partial_z\psi_n\) remain continuous at each
interface, and adjacent-field overlaps preserve state identity
\cite{ando1982,bendaniel1966}.

For normalized real envelope functions, the transition quantities are
\begin{align}
 E_{12}(F)&=E_2(F)-E_1(F)
\label{eq:E12}\\
 z_{12}(F)&=\int \psi_1(z;F)\,z\,\psi_2(z;F)\,dz
\label{eq:z12}\\
 f_{12}(F)&=
 \frac{2m_{12}^{\ast}(F)E_{12}(F)}{\hbar^2}
 |z_{12}(F)|^2
\label{eq:f12}
\end{align}
The transition reference mass \(m_{12}^{\ast}\) in Eq.~(\ref{eq:f12}) is
distinct from \(m^\ast(z,E_n)\). The sweep contains 21 fields from \(-100\)
to \(+100~\mathrm{kV\,cm^{-1}}\) in \(10~\mathrm{kV\,cm^{-1}}\) steps. The
resulting quantum quantities enter the resonant \(zz\) component of the MQW
permittivity. An intersubband transition contributes primarily to this tensor
component, for which a conventional Lorentz representation is
\begin{equation}
\chi_{zz}(\omega,F)=
\frac{N_v(F)e^2 f_{12}(F)}
{\epsilon_0 m_{12}^{\ast}(F)}
\frac{1}
{\omega_{12}^2(F)-\omega^2-i\Gamma_{12}^{\mathrm{L}}(F)\omega}
\label{eq:chi}
\end{equation}
where \(N_v=[N_1-N_2]/(23.7~\mathrm{nm})\),
\(\omega_{12}=E_{12}/\hbar\), and the Lorentz damping coefficient is
\(\Gamma_{12}^{\mathrm{L}}=2\gamma_{12}/\hbar\). Thus \(\Gamma_{12}^{\mathrm{L}}\)
has unit \(\mathrm{s^{-1}}\), whereas \(\gamma_{12}\) below is an energy
HWHM in meV. The exact \(23.7~\mathrm{nm}\) microscopic period normalizes
the population difference, and all energies, lengths, densities, and masses
are converted to SI units before Eq.~(\ref{eq:chi}) is evaluated
\cite{ando1982,helm2000}. RCWA uses this field-resolved anisotropic tensor at
normal incidence and \(x\) polarization; the TCMT comparison uses the common
\(5.0\)--\(7.6~\mu\mathrm{m}\) window. Retaining complex \(S_{11}\) is
essential because reflectance alone does not determine phase or off-axis
zero topology.
Equation~(\ref{eq:chi}) is dimensionally closed: its first factor has units
of \(\mathrm{s^{-2}}\), the Lorentz denominator has the same units, and
\(\chi_{zz}\) is dimensionless. The background dielectric response is added
to \(\chi_{zz}\) to form the relative-permittivity tensor used by RCWA. No
near-field observable is fitted. Neighboring-field eigenstate overlaps are
used only to preserve state identity through the QCSE sweep. All 21
self-consistent solutions satisfy the \(0.001~\mathrm{meV}\) potential-update
tolerance, and the adjacent-field state fidelity remains above 0.9997.

\section{One-port temporal coupled-mode theory}

RCWA is the reference electromagnetic solution and TCMT its constrained
reduced representation. With modal-energy normalization for the cavity and
ISBT amplitudes \(a\) and \(b\), power normalization for \(s_\pm\), and
\(\exp(-i\omega t)\) time dependence, the one-port equations are
\begin{align}
\frac{da}{dt}
&=
\left(-i\omega_{\mathrm{pc}}-\kappa_{\mathrm{pc}}\right)a
+ig_\omega(F)b+\sqrt{2\kappa_e}\,s_+
\label{eq:tcmt_time_a}\\
\frac{db}{dt}
&=
\left[-i\omega_{12}(F)-\kappa_{12}(F)\right]b
+ig_\omega(F)a
\label{eq:tcmt_time_b}\\
s_-&=-s_+ + \sqrt{2\kappa_e}\,a
\label{eq:io}
\end{align}
Only the bright cavity is driven directly. The background minus sign is the
metal-backed reflection phase. We use
\(E_j=\hbar\omega_j\), \(\gamma_j=\hbar\kappa_j\), and
\(g_{\mathrm{fit}}=\hbar g_\omega\), with
\(\gamma_{\mathrm{pc}}=\gamma_e+\gamma_{\mathrm{pc},\mathrm{nr}}\). Every
\(\gamma\) is an amplitude HWHM energy in meV. The matter rate
\(\gamma_{12}\) describes optical coherence loss, not population decay, and
is distinct from \(\Gamma_{12}^{\mathrm{L}}\). With this normalization and
time convention, steady-state elimination of the matter amplitude gives the
dressed-cavity denominator below \cite{fan2003,suh2004,ruan2010}
\begin{align}
D(E,F)={}&\gamma_{\mathrm{pc}}-i(E-E_{\mathrm{pc}})
\nonumber\\
&+\frac{g_{\mathrm{fit}}^2(F)}
{\gamma_{12}(F)-i[E-E_{12}(F)]}
\label{eq:D}
\end{align}
the reflection coefficient is
\begin{equation}
r(E,F)=-1+\frac{2\gamma_e}{D(E,F)}
\label{eq:r}
\end{equation}
The modeled reflectance is \(|r|^2\). With
\(\Delta(F)=E_{12}(F)-E_{\mathrm{pc}}\), the two internal polariton poles are
\begin{align}
\widetilde E_\pm(F)
={}&
\frac{E_{\mathrm{pc}}+E_{12}(F)}{2}
-i\frac{\gamma_{\mathrm{pc}}+\gamma_{12}(F)}{2}
\nonumber\\
&\pm
\sqrt{
g_{\mathrm{fit}}^2(F)+
\frac{
\left[\Delta(F)+i(\gamma_{\mathrm{pc}}-\gamma_{12}(F))\right]^2
}{4}
}
\label{eq:poles}
\end{align}
Their real parts are the LP and UP energies and their negative imaginary
parts the amplitude HWHM values. The two branches are joint eigenmodes of one
cavity--ISBT coupling, not independent channels. Pole-splitting and
linewidth criteria follow directly from the complex pole discriminant
\cite{anappara2009,geiser2012,benz2013}. The
matter term in Eq.~(\ref{eq:D}) is a frequency-dependent self-energy: its
real and imaginary parts shift and broaden the bright photonic response.
Accordingly, \(E_{\mathrm{pc}}\) is the effective bare photonic energy of the
two-mode reduction, not the independently extracted intersubband-off cavity
minimum \(E_{\mathrm{bc}}\). At exact resonance, distinct real parts of the
poles require
\(g_{\mathrm{fit}}> |\gamma_{\mathrm{pc}}-\gamma_{12}|/2\); a more
conservative linewidth-resolvability benchmark replaces the difference by
the sum of the two HWHM values. Along the nominal field trajectory, the
minimum sum-linewidth ratio is 0.928 and the minimum cooperativity
\({\cal C}=g_{\mathrm{fit}}^2/(\gamma_{\mathrm{pc}}\gamma_{12})\) is 0.937.
The system is therefore kept near a linewidth-dependent boundary rather than
classified from a visual anticrossing alone.

The
external reflection zeros instead satisfy \(r=0\), or
\begin{equation}
\begin{split}
&\left[E-E_{\mathrm{pc}}
+i(\gamma_{\mathrm{pc},\mathrm{nr}}-\gamma_e)\right]\\
&\qquad\times
\left[E-E_{12}(F)+i\gamma_{12}(F)\right]-g_{\mathrm{fit}}^2(F)=0
\end{split}
\label{eq:zero_polynomial}
\end{equation}
For \(E_{z,j}=\operatorname{Re}E_{z,j}+i\operatorname{Im}E_{z,j}\) and the
adopted time convention, the external regimes are
\begin{equation}
\begin{aligned}
\operatorname{Im}E_{z,j}<0
&\quad &&\text{undercoupled}\\
\operatorname{Im}E_{z,j}=0
&&&\text{critically coupled}\\
\operatorname{Im}E_{z,j}>0
&&&\text{overcoupled}
\end{aligned}
\qquad j=\mathrm{LP},\mathrm{UP}
\label{eq:zero_regimes}
\end{equation}
Critical coupling is therefore a real-axis zero crossing and is distinct from
internal strong coupling \cite{krasnok2019}. The zero equation contains both
the direct reflection amplitude and the radiative--nonradiative balance. A
deep real-frequency reflectance minimum is consequently suggestive, but does
not by itself prove that a complex zero lies on the real axis. At simultaneous
cavity--matter resonance the zero condition reduces to
\(\gamma_e=\gamma_{\mathrm{pc},\mathrm{nr}}+
g_{\mathrm{fit}}^2/\gamma_{12}\); away from this special point, the full
complex roots of Eq.~(\ref{eq:zero_polynomial}) must be followed.

A microscopic post-fit benchmark
follows from the susceptibility strength. Defining
\(\mathcal{S}(F)=[N_1(F)-N_2(F)]f_{12}(F)/m_{12}^{\ast}(F)\), and denoting by
\(\eta_S(F)\) the normalized distributed polarization--field overlap after
removing the dipole-strength trend already contained in \(\mathcal{S}\),
\begin{equation}
g_{\mathrm{mic}}(F)=g_{\mathrm{fit}}(F_{\mathrm{ref}})
\left[\frac{\mathcal{S}(F)}{\mathcal{S}(F_{\mathrm{ref}})}\right]^{1/2}
\eta_S(F)
\label{eq:g_scaling}
\end{equation}
with \(F_{\mathrm{ref}}=0\). The relative strength
\(\mathcal{S}=[N_1-N_2]f_{12}/m_{12}^{\ast}\) carries the population,
oscillator-strength, and mass dependence; \(\eta_S\) is a normalized spatial
overlap audit and is not fitted. Thus \(g_{\mathrm{mic}}\) tests whether the
field trend of the fitted effective coupling is compatible with the
microscopic oscillator-strength and mode-overlap trend; it is not substituted
for \(g_{\mathrm{fit}}\) in the pole--zero result. Residue and overlap
normalizations are evaluated in SI units. The Lorentz residue reconstructed
from the S--P transition strength agrees with the optical-tensor residue to a
maximum relative error of \(8.2\times10^{-14}\), while the distributed-overlap
correction changes the coupling by less than
\(3\times10^{-4}~\mathrm{meV}\) \cite{todorov2012,todorov2015}.

The zero-field complex fit determines
\(E_{\mathrm{pc}}\), \(g_{\mathrm{fit}}(0)\), \(\gamma_e\),
\(\gamma_{\mathrm{pc},\mathrm{nr}}\), and a common reference-plane phase
\begin{equation}
\vartheta(E)=\phi_c+\alpha(E-E_{\mathrm{ref}})
\label{eq:phase_delay}
\end{equation}
where \(E_{\mathrm{ref}}=200~\mathrm{meV}\). The fitted
\(\alpha=0.07118^\circ\,\mathrm{meV}^{-1}\) equals
\(1.24231\times10^{-3}~\mathrm{rad\,meV^{-1}}\) and corresponds to
\(\tau_{\mathrm{ref}}=\hbar\alpha=0.8177~\mathrm{fs}\). The phase parameters
are converted to radians in the exponential and fixed after calibration. At
the remaining fields, S--P fixes \(E_{12}\), the dephasing model fixes
\(\gamma_{12}\), and only \(g_{\mathrm{fit}}(F)\) is fitted. The objective
minimizes the complex-amplitude residual, not reflectance alone. One hundred
randomized zero-field initializations probe uniqueness, while the positive-
and negative-field sequences are continued independently from zero using the
neighboring solution. Repeating the fit over four wavelength windows tests
its stability. This separation between shared,
microscopically prescribed, and singly fitted field-dependent quantities is
what makes the extracted zero trajectories identifiable.
The parameter provenance is therefore compact: the photonic and phase
baseline is shared, \(E_{12}(F)\) and \(\gamma_{12}(F)\) are prescribed by
the quantum and dephasing models, and \(g_{\mathrm{fit}}(F)\) is the sole
field-dependent fitted quantity. Energies and rates are amplitude HWHM
quantities in meV, phase parameters are converted to radians before complex
exponentials, and every parameter is traceable to one of these three source
classes.

\section{Electrical control of polariton states and critical coupling}

The tracked transition in Fig.~\ref{fig:qcse}(a) shifts from 173.821 to
\(221.523~\mathrm{meV}\), spanning \(47.702~\mathrm{meV}\) and crossing the
shared fitted cavity energy. At zero field,
\(E_{12}=199.282~\mathrm{meV}\), \(z_{12}=1.5420~\mathrm{nm}\), and
\(f_{12}=0.5936\). All fields converge with adjacent-field fidelity above
0.9997.

\begin{plainwidetext}
\centering
\includegraphics[width=\textwidth]{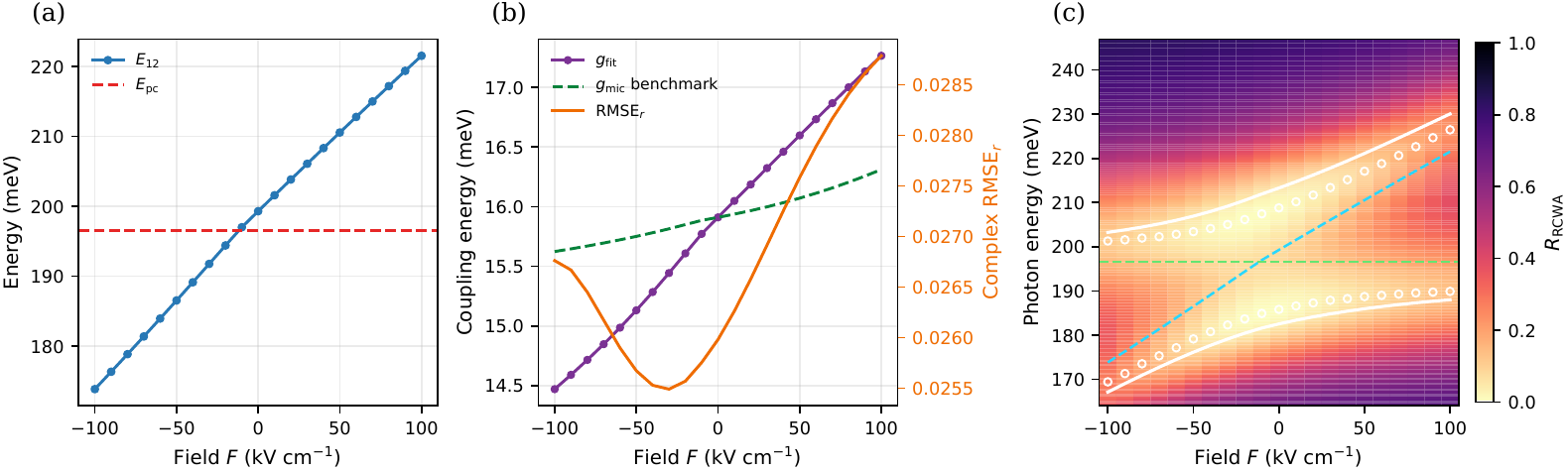}
\widefigurecaption{Field-resolved QCSE, RCWA, and reduced-model summary.
(a) Tracked \(E_{12}(F)\) and the shared fitted \(E_{\mathrm{pc}}\).
(b) Effective coupling extracted from the complex far field, the
zero-field-normalized microscopic strength-and-overlap benchmark of
Eq.~(\ref{eq:g_scaling}), and the
complex-amplitude RMSE. (c) Field-resolved RCWA reflectance
\(R_{\mathrm{RCWA}}(E,F)\). White solid lines are the real parts of the TCMT
poles, dashed lines are the bare matter and fitted cavity energies, and open
circles are the directly extracted RCWA minima.}
\label{fig:qcse}
\end{plainwidetext}

Fig.~\ref{fig:qcse}(b) shows the central reduction. The zero-field-normalized
microscopic benchmark varies from 15.62 to \(16.31~\mathrm{meV}\), whereas
the complex far field requires a smooth
\(g_{\mathrm{fit}}=14.470\)--\(17.264~\mathrm{meV}\). The two trajectories
give similar aggregate reflectance RMSE values, 0.02923 and 0.02915,
respectively, but the microscopic trajectory raises the complex-amplitude
RMSE from 0.02680 to 0.03875. Because
both are anchored at zero field, this difference isolates effective
field-dependence rather than the absolute coupling scale. The overlap
correction is below \(3\times10^{-4}~\mathrm{meV}\).
The remaining difference is physically plausible because a far-field
two-mode coupling absorbs radiative mode reshaping and background-reduction
errors that are not contained in a scalar oscillator-strength law. The
agreement in scale supports the microscopic interpretation, while the lower
complex RMSE justifies using \(g_{\mathrm{fit}}\) for the quantitative zero
trajectories.

The RCWA minima in Fig.~\ref{fig:qcse}(c) approach, avoid crossing, and
separate. Their offset from the real parts of the poles is physical because a
lossy one-port scattering extremum need not coincide with an internal pole.
The linewidth ratio and cooperativity place the system near a
linewidth-dependent boundary. Scaling the complete \(\gamma_{12}(F)\)
trajectory by 0.8 and 1.2 preserves the qualitative anticrossing but shifts
the interpolated LP critical field from \(-12.96\) to
\(-4.08~\mathrm{kV\,cm^{-1}}\) and the UP field from \(-11.11\) to
\(-17.63~\mathrm{kV\,cm^{-1}}\); the boundary is therefore not used as a
binary claim.
Specifically, the real pole separation is controlled by the complex
eigenvalue discriminant, while the observed dip separation is additionally
shifted by the direct reflection path and by the location of the scattering
zeros. The anticrossing is therefore evidence for hybridization, but its
visual gap is not an independent measurement of \(2g_{\mathrm{fit}}\).

Fig.~\ref{fig:single} tests the zero-field reduction over
\(5.0\)--\(7.6~\mu\mathrm{m}\). Fig.~\ref{fig:single}(a) provides the
phase-sensitive test: one parameter set follows both quadratures of the
complex trajectory. Fig.~\ref{fig:single}(b) shows that the same parameters
reproduce the UP and LP reflectance minima at \(5.938\) and
\(6.672~\mu\mathrm{m}\). The smooth residual in Fig.~\ref{fig:single}(c)
contains no additional resonant feature, supporting the adequacy of the
two-mode response within the fitting window. Across the sweep the two minima
remain continuous and reach a minimum separation of \(22.482~\mathrm{meV}\)
near zero detuning. An intersubband-off RCWA check gives
\(E_{\mathrm{bc}}=196.457~\mathrm{meV}\), only \(0.091~\mathrm{meV}\) below
\(E_{\mathrm{pc}}\). At this energy, the calculated \(|E_z|\) profile is
confined to the \(750~\mathrm{nm}\)-wide MQW-loaded resonator and is used only
as a post-fit validation \cite{gabbay2012,campione2014}.
This agreement is an independent consistency check rather than an equality
constraint: \(E_{\mathrm{bc}}\) is a minimum of a lossy full-wave spectrum,
whereas \(E_{\mathrm{pc}}\) is the bare complex-mode parameter inferred from
the joint response.

\begin{plainwidetext}
\centering
\includegraphics[width=\textwidth]{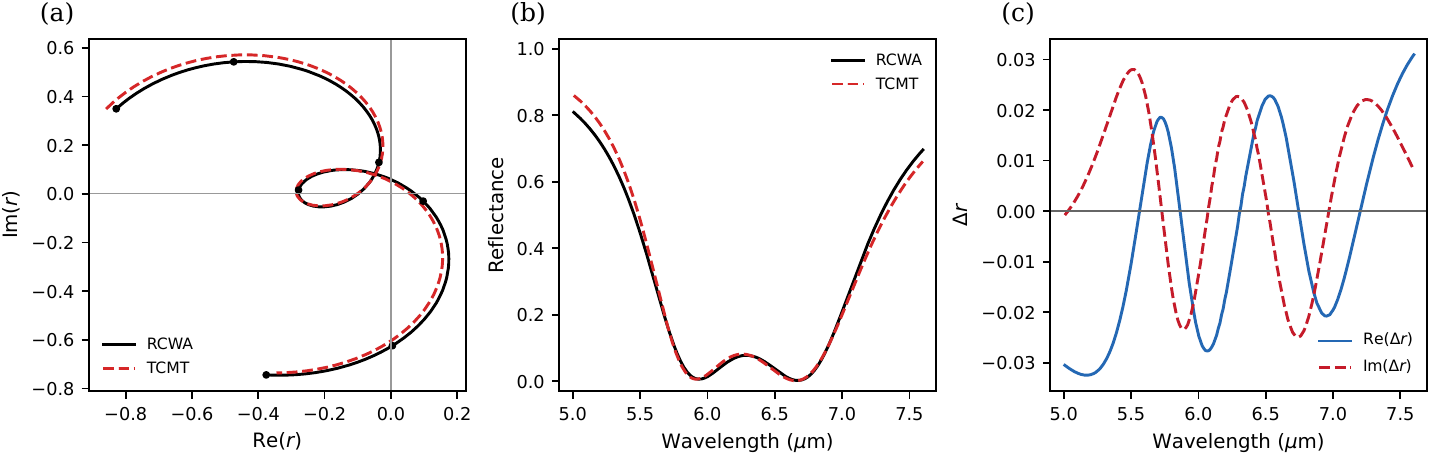}
\widefigurecaption{Zero-field complex-response comparison.
(a) Full-wave RCWA coefficient \(S_{11}\) and the aligned one-port TCMT
coefficient in the complex plane. (b) Reflectance spectra generated by the
same parameter set. (c) Complex residual
\(\Delta r=e^{i\vartheta(E)}r_{\mathrm{TCMT}}-S_{11}^{\mathrm{RCWA}}\).
The common phase baseline is
\(\phi_c=10.759^\circ\) and
\(\alpha=0.0712^\circ\,\mathrm{meV}^{-1}\), corresponding to
\(\tau_{\mathrm{ref}}=0.818~\mathrm{fs}\).}
\label{fig:single}
\end{plainwidetext}

A reflectance-only control lowers \(\mathrm{RMSE}_R\) from 0.0280 to 0.0181
but gives \(\mathrm{RMSE}_r=0.0562\), more than twice the complex-fit value
0.0260, while shifting the fitted rates. Thus intensity does not uniquely fix
the zero topology. All 100 randomized initializations recover the same
baseline solution, and changing the fitting window changes
\(g_{\mathrm{fit}}(0)\) by at most \(0.043~\mathrm{meV}\).
The control demonstrates why intensity-only fitting is inadequate here: two
models can reproduce nearly the same dip depths while assigning different
radiative rates and hence different off-axis zeros. Complex calibration
retains both quadratures of the scattering coefficient and constrains that
ambiguity.

\FloatBarrier

\begin{figure*}[t]
\centering
\includegraphics[width=\textwidth]{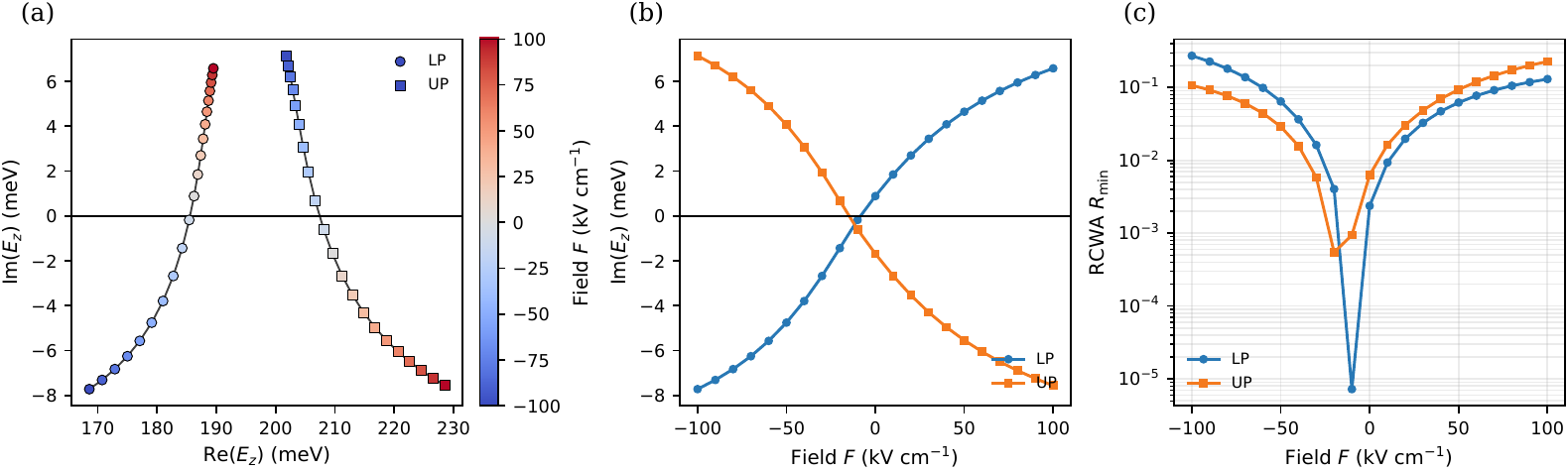}
\caption{Field evolution of the polaritonic reflection zeros and a full-wave
real-axis-zero cross-check. (a) Complex-energy trajectories of the lower-polariton (LP,
circles) and upper-polariton (UP, squares) zeros obtained from the 21
constrained TCMT fits; marker color denotes the applied field. The horizontal
line is the real-energy axis. (b) Imaginary parts of the two zero energies
versus field; positive and negative values denote overcoupled and
undercoupled one-port conditions, respectively. (c) Directly sampled RCWA
reflectance minima. The robust crossing brackets are
\([-10,0]~\mathrm{kV\,cm^{-1}}\) for LP and
\([-20,-10]~\mathrm{kV\,cm^{-1}}\) for UP in both descriptions.}
\label{fig:zeros}
\end{figure*}

The fitted \(E_{\mathrm{pc}}=196.548~\mathrm{meV}\) lies inside the QCSE
range, and \(E_{12}=E_{\mathrm{pc}}\) occurs between \(-20\) and
\(-10~\mathrm{kV\,cm^{-1}}\). The two minima avoid crossing, but their
separation is not \(2g_{\mathrm{fit}}\) because damping, the direct path, and
the pole--zero distinction shift intensity extrema. The same constrained
parameters generate all 42 reflection zeros in Fig.~\ref{fig:zeros}.
Branch labels are assigned continuously from their limiting matter- and
photon-like character rather than re-sorted independently at each field.
This prevents an apparent LP--UP exchange from being introduced by root
ordering. The same continuation convention is used for poles, zeros, RCWA
minima, and overlap-tracked quantum states.

The continuous complex-plane paths in Fig.~\ref{fig:zeros}(a) establish the
branch identities and show that critical coupling is reached by motion of a
scattering zero, not merely by a reduction in dip depth. Each branch crosses
the real axis once [Fig.~\ref{fig:zeros}(b)]. With
increasing field, the LP zero changes from undercoupled to overcoupled, while
the UP zero crosses in the opposite direction. The robust brackets are
\([-10,0]~\mathrm{kV\,cm^{-1}}\) for LP and
\([-20,-10]~\mathrm{kV\,cm^{-1}}\) for UP; TCMT interpolation gives
\(-8.28\) and \(-14.73~\mathrm{kV\,cm^{-1}}\). This branch-selective
redistribution of radiative and dissipative balance is external and remains
distinct from internal strong coupling.
The QCSE acts on several quantities simultaneously: it sweeps the transition
through the cavity, changes oscillator strength and dephasing, and produces a
smooth effective-coupling correction. These changes alter the photonic and
matter fractions of both branches. Their opposite zero-crossing directions
therefore represent a transfer of the one-port radiative--dissipative balance
between the two hybrid modes, not two independently tuned absorbers.

At \(-10~\mathrm{kV\,cm^{-1}}\), the sampled-grid LP and UP RCWA minima are
\(5.66\times10^{-5}\) and \(9.70\times10^{-4}\), respectively. Solving
\(\operatorname{Re}S_{11}=\operatorname{Im}S_{11}=0\) by bilinear
interpolation of the raw complex grid places both full-wave crossings in the
same sampled-field brackets as TCMT [Fig.~\ref{fig:zeros}(c)]. The
interpolated RCWA fields, \(-9.41\) and
\(-16.00~\mathrm{kV\,cm^{-1}}\), differ from TCMT by 1.13 and
\(1.27~\mathrm{kV\,cm^{-1}}\); the brackets are therefore the robust result.
The sub-grid values depend on interpolation and the adopted dephasing model,
whereas the sign changes at sampled fields do not. We therefore report the
brackets as the principal prediction and retain decimal critical fields only
as model-conditioned estimates.
The agreement criterion is therefore topological and grid-resolved: TCMT and
full-wave RCWA must place the sign change of each complex reflection amplitude
in the same adjacent-field interval. Agreement of two interpolated decimal
fields would be a stronger numerical statement than the available sampling
supports and is not required for the reported conclusion.
Experimental MQW--plasmonic devices have already demonstrated electrical
reflection-phase control on this material and device platform
\cite{chung2023}. The present result extends that experimental foundation to
a testable prediction of branch-resolved complex-zero crossings rather than
claiming their direct experimental observation.

\section{Conclusion}

We established a field-consistent multiscale reduction of an electrically
tuned intersubband-polaritonic metasurface. S--P supplies the quantum
trajectory, complex RCWA supplies the reference scattering amplitude, and one
zero-field calibration fixes the shared photonic and phase-delay baseline;
only \(g_{\mathrm{fit}}(F)\) varies in the remaining fits. The model maintains
\(\mathrm{RMSE}_r=0.0255\)--0.0288, while independent bare-cavity, residue,
overlap, multistart, and window controls constrain its interpretation. Most
importantly, internal poles and external reflection zeros respond differently
to the QCSE: the LP and UP zeros cross the real axis in opposite directions,
and TCMT and direct complex RCWA recover the same two critical-coupling field
intervals. The sub-grid fields remain dephasing- and interpolation-dependent,
whereas the branch topology and sampled-field brackets are robust. This
identifiable pole--zero construction is transferable to electrically tuned
polaritonic systems with a common photonic baseline.

\begin{acknowledgments}
The author conducted this research independently, without institutional
affiliation or external financial support.
\end{acknowledgments}

\section*{Data availability}
The data that support the findings of this study are available from the
corresponding author upon reasonable request.

\section*{References}
\makeatletter
\renewcommand{\bibsection}{}
\makeatother
\bibliography{references}

\onecolumngrid
\clearpage
\twocolumngrid
\clearpage
\hypersetup{pageanchor=false}
\setcounter{page}{1}
\setcounter{section}{0}
\setcounter{figure}{0}
\setcounter{table}{0}
\setcounter{equation}{0}
\renewcommand{\thefigure}{S\arabic{figure}}
\renewcommand{\thetable}{S\arabic{table}}
\renewcommand{\theequation}{S\arabic{equation}}
\renewcommand{\theHfigure}{S\arabic{figure}}
\renewcommand{\theHtable}{S\arabic{table}}
\renewcommand{\theHequation}{S\arabic{equation}}
\let\maketitle\savedapsmaketitle
\let\author\savedapsauthor
\let\email\savedapsemail
\let\affiliation\savedapsaffiliation
\pagestyle{footerpage}

\makeatletter
\renewcommand{\section}{%
  \@startsection
    {section}%
    {1}%
    {\z@}%
    {1.0cm}%
    {0.5cm}%
    {\normalfont\small\bfseries\centering}%
}
\makeatother

\title{Supplemental Material for ``Multiscale Analysis of Electrically Tunable Reflection Zeros in Intersubband Polaritonic Metasurfaces''}
\author{Inyong Hwang}
\email{makki5068@gmail.com}
\affiliation{Independent Researcher, Hwaseong-si, Gyeonggi-do, Republic of Korea}
\date{August 17, 2026}
\makeatletter
\let\apscombinedlabel\label
\def\label#1{%
  \def\apscombinedtempa{#1}%
  \def\apscombinedtempb{FirstPage}%
  \ifx\apscombinedtempa\apscombinedtempb\else\apscombinedlabel{#1}\fi}
\maketitle
\let\label\apscombinedlabel
\makeatother
\thispagestyle{footerpage}

\section{Quantum-well and RCWA numerical settings}

The asymmetric double quantum well used for the field sweep is summarized in
Table~\ref{tab:layers}. The S--P domain contains one \(23.7~\mathrm{nm}\)
unit. For the electromagnetic geometry, this microscopic period was assigned
a nominal thickness of \(25~\mathrm{nm}\) and repeated eight times, giving
the \(200~\mathrm{nm}\) homogenized MQW layer used in RCWA. The exact
unrounded thickness is \(8\times23.7=189.6~\mathrm{nm}\); the rounding is
applied only to the RCWA geometric thickness. In the underlying microscopic
sequence, the two terminal \(8~\mathrm{nm}\) barriers form the intended
\(16~\mathrm{nm}\) separation barrier between neighboring asymmetric double
wells. The S--P calculation resolves the exact local period that defines the
nominal eight-period electromagnetic layer.

\begin{table}[b]
\caption{Layer sequence used in the S--P calculation.}
\label{tab:layers}
\resizebox{\columnwidth}{!}{%
\begin{tabular}{lccc}
\toprule
Layer & Material & Thickness (nm) & Doping \\
\midrule
Left barrier & InAlAs & 8.0 & undoped\\
Wide well & InGaAs & 5.0 & \(5\times10^{18}~\mathrm{cm^{-3}}\) Si\\
Tunnel barrier & InAlAs & 1.0 & undoped\\
Narrow well & InGaAs & 1.7 & undoped\\
Right barrier & InAlAs & 8.0 & undoped\\
\bottomrule
\end{tabular}
}
\end{table}

The substrate lattice reference is InP with \(a=5.8687~\text{\AA}\).
The material parameters used by the solver are
\(m_e=0.0840m_0\), \(\epsilon_r=12.707\) for InAlAs, and
\(m_e=0.0437m_0\), \(\epsilon_r=14.093\) for InGaAs before the
energy-dependent mass correction. The conduction-band offset magnitude
between the two materials is \(514.424~\mathrm{meV}\).
The Hamiltonian uses the BenDaniel--Duke ordering
\(-(\hbar^2/2)\partial_z[m^{\ast-1}(z,E)\partial_z]\), so the distinct
InGaAs and InAlAs masses are retained and the normal probability current is
continuous at every interface.

The nominal spatial step is \(0.1~\text{\AA}\), giving 2371 points over the
\(23.7~\mathrm{nm}\) domain; the dense eigenvalue step uses the factor-of-two
downsampling recorded in the solver export. Hard-wall envelope conditions
\(\psi_n(0)=\psi_n(23.7~\mathrm{nm})=0\) and zero-reference Dirichlet
conditions for the Hartree potential at the two domain edges are used. The
self-consistent cycle uses linear potential mixing with \(\beta=0.08\), at
most 200 iterations, and a \(0.001~\mathrm{meV}\) maximum-update stopping
criterion. The source solver evaluates the subband populations with the
two-dimensional Fermi--Dirac expression at \(T=300~\mathrm{K}\). It applies
the Kane-type nonparabolic correction
\(E_{\mathrm{np}}(1+\alpha E_{\mathrm{np}})=E_{\mathrm{p}}\), with
\(\alpha=0.20/E_{g,\mathrm{eff}}\) and a wavefunction-weighted effective
bandgap. The field-resolved masses and subband populations exported by this
implementation are used directly below.

The signed field parameter follows the numerical convention
\[
U_{\mathrm{ext}}(z)=eF(z-z_c)
\]
where \(z_c\) is the center of the local quantum domain. Consequently,
positive \(F\) corresponds to the physical electric field
\(\mathbf{E}_{\mathrm{ext}}=F\hat{\mathbf z}\). This convention fixes the
sign of the field axis; the additive reference-energy shift does not affect
the reported transition energies.

The Hartree potential entering the main-text Hamiltonian is updated from
\begin{equation}
\frac{d}{dz}\left[
\epsilon_0\epsilon_r(z)\frac{d\phi}{dz}
\right]=-\rho(z)
\qquad
\rho(z)=e[N_D^+(z)-n(z)]
\label{eq:poisson_s}
\end{equation}
where \(e>0\), \(N_D^+\) is the ionized-donor density, and \(n\) is the
occupied-electron density. The electron-energy contribution is \(-e\phi\),
so Eqs.~(\ref{eq:poisson_s}) and the main-text
\(U_{\mathrm{ext}}=eF(z-z_c)\) convention fix all electrostatic signs. At an
abrupt heterointerface, the BenDaniel--Duke operator imposes continuity of
both \(\psi_n\) and
\(m^{\ast-1}\partial\psi_n/\partial z\), conserving the normal probability
current. The normalized \(\psi_n\) are conduction-band envelope functions;
the transition reference mass \(m_{12}^{\ast}\) used in the oscillator
strength is distinct from the local energy-dependent mass \(m^\ast(z,E_n)\).

The donor sheet density is specified per local period. If \(N_1(F)\) and
\(N_2(F)\) are the calculated subband sheet populations, the optical tensor
uses
\[
N_v(F)=\frac{N_1(F)-N_2(F)}{23.7~\mathrm{nm}}
=\frac{8[N_1(F)-N_2(F)]}{189.6~\mathrm{nm}}
\]
The calculated subband populations include the small upper-subband
occupation through the field-resolved difference \(N_1(F)-N_2(F)\). This
population-density normalization retains the exact microscopic period;
\(200~\mathrm{nm}\) is the nominal geometric thickness assigned to the
homogenized RCWA layer.

All dimensional substitutions in the susceptibility and residue calculations
are performed in SI units: energies are converted from meV to J, lengths from
nm to m, sheet densities from \(\mathrm{cm^{-2}}\) to \(\mathrm{m^{-2}}\), and
the tabulated transition-mass ratio is multiplied by \(m_0\) to obtain kg.
Presentation units are converted only after the calculation. When an
energy-domain residue is reported in \(\mathrm{eV^2}\), the equivalent
frequency-domain residue is multiplied by \(\hbar^2\) with \(\hbar\) expressed
in \(\mathrm{eV\,s}\).

To connect these quantum inputs to the optical response, the generating RCWA
configuration uses a \(4.000\)--\(10.000~\mu\mathrm{m}\)
grid of 501 uniformly spaced points, corresponding to
\(0.012~\mu\mathrm{m}\) increments. The computational periods are
\(1.1~\mu\mathrm{m}\) along both in-plane axes; the structure is invariant
along \(y\). The centered top-Au and MQW rectangles both have an
\(x\)-span of \(0.75~\mu\mathrm{m}\), while the continuous bottom-Au layer
spans the full \(1.1~\mu\mathrm{m}\) unit cell. Their thicknesses are
\(0.05\), \(0.20\), and \(0.20~\mu\mathrm{m}\), respectively. The Fourier
truncation orders are \(N_x=N_y=7\). The calculation is performed at normal
incidence and zero azimuth with the linear \(x\)-polarized, TM-like channel,
retaining the forward \(S_{11}\) and \(S_{21}\) coefficients. Vacuum is the
incident medium, the output substrate uses the tabulated Si response, the Au
layers use the corrected mid-infrared Drude data, and the MQW layer uses the
field-resolved diagonal-anisotropic tensor.

For the TCMT reduction, every one of the 21 field-resolved complex spectra is
restricted to 217 wavelength samples from \(5.008\) to
\(7.600~\mu\mathrm{m}\), retaining the same \(0.012~\mu\mathrm{m}\)
spacing. The fit uses
the concatenated real and imaginary residuals of \(S_{11}\), without an
additional wavelength-dependent weight. The zero-field fit is performed with
the trust-region least-squares implementation in SciPy. The bounds are
160--240 meV for \(E_{\mathrm{pc}}\), 0--50 meV for \(g_{\mathrm{fit}}(0)\), \(\gamma_e\),
and \(\gamma_{\mathrm{pc},\mathrm{nr}}\), \(-180^\circ\)--\(180^\circ\)
for \(\phi_c\), and \(-2\)--\(2^\circ\,\mathrm{meV}^{-1}\) for \(\alpha\).
The termination tolerances are
\(\mathrm{xtol}=\mathrm{ftol}=\mathrm{gtol}=10^{-13}\), with at most 30000
function evaluations. At nonzero field only \(g_{\mathrm{fit}}(F)\) is fitted, within
0--50 meV, with at most 10000 evaluations and the same tolerances.

The archived RCWA exports retain the complex scattering coefficients,
wavelength grid, and energy-balance diagnostics, while the full-settings JSON
records the generating geometry and Fourier truncation. As numerical checks
on the exported solutions, the maximum transmission over the complete
field--wavelength grid is \(6.68\times10^{-9}\) and the maximum absolute
energy-balance residual is below \(3\times10^{-16}\). The accompanying data and
reduced outputs support an audit of the complete downstream TCMT reduction
from these archived complex spectra.

\section{Self-consistency and state tracking}

\begin{figure}[t]
\centering
\includegraphics[width=\columnwidth]{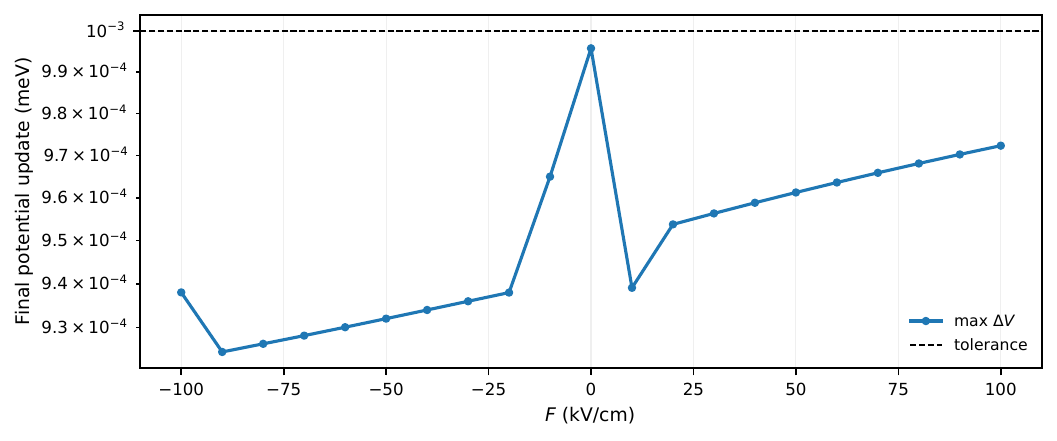}
\caption{Final maximum potential update at each applied field and the
prescribed \(0.001~\mathrm{meV}\) tolerance.}
\label{fig:sconv}
\end{figure}

\begin{figure}[t]
\centering
\includegraphics[width=\columnwidth]{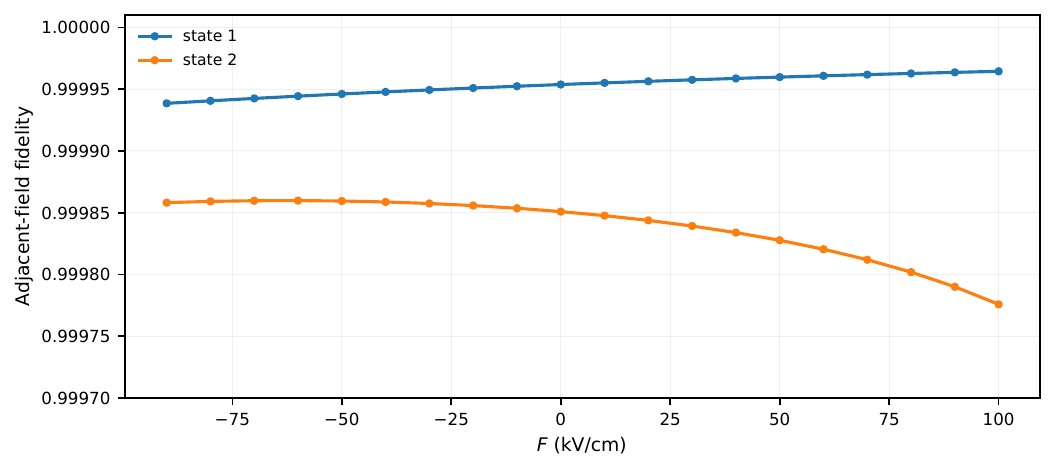}
\caption{Adjacent-field fidelity for the two overlap-tracked states. The
energy-order label is unchanged throughout the sweep.}
\label{fig:tracking}
\end{figure}

The final potential update at every field is below the stopping tolerance
[Fig.~\ref{fig:sconv}]. The overlap tracker associates every state with the
preceding field point, and the energy-order label and phase remain continuous.
Fig.~\ref{fig:tracking}
shows that the adjacent-field fidelity exceeds 0.9997 for both states. All
calculated wave functions are normalized.

\section{Optical dephasing model}

In the absence of a measured field-resolved linewidth for the modeled
heterostructure, the optical calculation uses the phenomenological dephasing
relation
\begin{equation}
W_{12}^{\mathrm{FWHM}}(F)=
\sqrt{(10~\mathrm{meV})^2+
(0.15E_F^{2\mathrm{D}})^2+
(0.04E_{12})^2}
\end{equation}
where \(E_F^{2\mathrm{D}}\) is the solver-derived two-dimensional Fermi
energy associated with the field-resolved subband population. The constant
10-meV term and the two dimensionless coefficients define the assumed
inhomogeneous, carrier-energy, and transition-energy contributions; they are
model inputs rather than independently predicted scattering rates. All
energies inside the square root are evaluated in meV, so
\(W_{12}^{\mathrm{FWHM}}\) is obtained in meV. Further,
\(\gamma_{12}(F)=W_{12}^{\mathrm{FWHM}}(F)/2\). The corresponding dephasing
time is \(T_2=\hbar/\gamma_{12}\); it is an optical dephasing time, not a
population lifetime. Here \(W_{12}^{\mathrm{FWHM}}\) and \(\gamma_{12}\) are,
respectively, the intensity FWHM and amplitude HWHM in energy units. The
nominal sweep gives \(\gamma_{12}=11.151\)--\(11.555~\mathrm{meV}\). This
scale is independently bracketed by the \(\gamma_{12}=8\), 10, and
\(14.3~\mathrm{meV}\) values examined for an experimentally realized
InGaAs/InAlAs intersubband-polaritonic metasurface based on the same MQW
platform~\cite{chung2023}; this comparison supports the linewidth scale, not
the phenomenological coefficients above. The
damping coefficient in the main-text second-order Lorentz denominator is the
distinct angular-frequency quantity
\(\Gamma_{12}^{\mathrm{L}}=
W_{12}^{\mathrm{FWHM}}/\hbar=2\gamma_{12}/\hbar\), with unit
\(\mathrm{s^{-1}}\).
The effective linewidth is therefore treated explicitly as a field-resolved
model input. Its influence is tested below by scaling the complete HWHM
trajectory by \(\pm20\%\) and repeating the full constrained calibration.
The resulting critical fields are consequently model-conditioned estimates;
for the nominal calibration, the sampled-field crossing intervals are more
robust than the sub-grid interpolation values. Their linewidth dependence is
tested below.

\section{Bare-cavity validation and microscopic-coupling benchmark}
\label{sec:microaudit}

\begin{figure}[t]
\centering
\includegraphics[width=0.92\columnwidth]{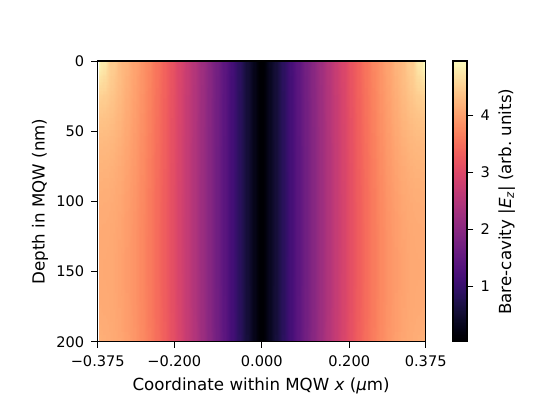}
\caption{Magnitude of the intersubband-off bare-cavity \(E_z\) field at
\(F=0\) and \(E_{\mathrm{bc}}=196.457~\mathrm{meV}\), restricted to the
\(750~\mathrm{nm}\)-wide patterned MQW region over the nominal
\(0\)--\(200~\mathrm{nm}\) RCWA depth. No metal or exterior region is shown.}
\label{fig:barefield}
\end{figure}

The intersubband-off RCWA calculation preserves the patterned cavity,
background tensor, metallic loss, and external port while removing the
resonant contribution from \(\epsilon_{zz}\). Its zero-field TM reflection
minimum occurs at
\(E_{\mathrm{bc}}=196.457~\mathrm{meV}\), within
\(0.091~\mathrm{meV}\) of the independently fitted effective photonic energy
\(E_{\mathrm{pc}}=196.548~\mathrm{meV}\). The complex field at
that minimum supplies the fixed driven profile in Fig.~\ref{fig:barefield}.
It is a scattering-normalized field, not an energy-normalized cavity
eigenmode, and is used only for the post-fit overlap audit below.

For real normalized envelope functions, the signed transition density is
\(\rho_{12}(z;F)=\psi_1(z;F)\psi_2(z;F)\). Following the electrical-dipole
gauge description of the collective intersubband polarization
\cite{todorov2012,todorov2015}, its normalized spatial envelope is
\begin{equation}
p_{12}(z;F)=
\frac{\displaystyle\int_0^z\rho_{12}(z';F)\,dz'}
{\displaystyle\left[\int\left|\int_0^z\rho_{12}(z';F)\,dz'\right|^2dz\right]^{1/2}}
\label{eq:p12_s}
\end{equation}
For microscopic period \(q\), the polarization--displacement-field matrix
element and distributed overlap are
\begin{align}
M_q(F,x)&=\int_{z_q}^{z_q+23.7\,\mathrm{nm}}
p_{12}(z-z_q;F)D_z(x,z)\,dz
\label{eq:mq_s}\\
S_{\mathrm{full}}(F)&=
\left[\sum_q\int_{\mathrm{active}}|M_q(F,x)|^2dx\right]^{1/2}
\label{eq:sfull_s}
\end{align}
where \(D_z=\epsilon_0\epsilon_{zz,\mathrm{bg}}E_z\). To avoid counting the
dipole-strength trend both in the susceptibility and the spatial integral,
the residual correction is
\begin{equation}
\eta_S(F)=
\frac{S_{\mathrm{full}}(F)/S_{\mathrm{full}}(0)}
{S_{\mathrm{dip}}(F)/S_{\mathrm{dip}}(0)}
\label{eq:eta_s}
\end{equation}
where \(S_{\mathrm{dip}}\) uses the corresponding dipole-approximation matter
profile.

\begin{figure*}[t]
\centering
\includegraphics[width=0.86\textwidth]{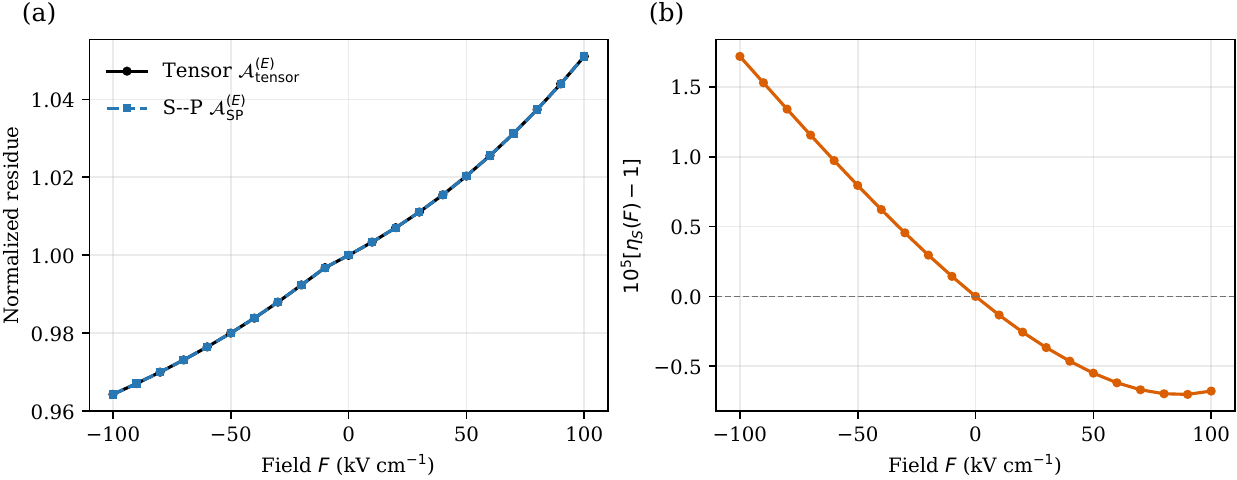}
\caption{Post-fit microscopic validation. (a) Population-difference-weighted
energy-domain Lorentz residue extracted from the corrected optical tensor and
reconstructed from the S--P transition strength, normalized at zero field.
(b) Residual distributed-overlap
correction after removing the dipole-strength trend. The displayed scale is
\(10^5[\eta_S(F)-1]\).}
\label{fig:microvalidation}
\end{figure*}

The normalization of the transition-strength component is checked separately
in the energy domain, matching the energy-squared residue stored in the
supporting data:
\begin{equation}
\chi_{zz}(E,F)=
\frac{\mathcal{A}_{\mathrm{tensor}}^{(E)}(F)}
{E_{12}^2(F)-E^2-i2\gamma_{12}(F)E}
\label{eq:lorentz_audit}
\end{equation}
and reconstructing its residue from the S--P output,
\begin{equation}
\mathcal{A}_{\mathrm{SP}}^{(E)}(F)=
\hbar^2
\frac{N_v(F)e^2f_{12}(F)}
{\epsilon_0m_{12}^{\ast}(F)}
\label{eq:asp_s}
\end{equation}
Here \(\mathcal{A}^{(E)}\) has unit energy squared; the tabulated data use
\(\mathrm{eV^2}\), with \(\hbar\) evaluated in \(\mathrm{eV\,s}\). All three
terms in the denominator of Eq.~(\ref{eq:lorentz_audit}) therefore have unit
\(\mathrm{eV^2}\), and \(\chi_{zz}\) is dimensionless. The actual
field-resolved density is
\(N_v(F)=[N_1(F)-N_2(F)]/L_p\), where \(L_p=23.7~\mathrm{nm}\).
The energy-domain residue extracted from the corrected RCWA tensor and the
S--P reconstruction agree to a maximum relative error of
\(8.2\times10^{-14}\) after CSV serialization over
all 21 fields [Fig.~\ref{fig:microvalidation}(a)]. The overlap correction
satisfies \(|\eta_S-1|<1.72\times10^{-5}\)
[Fig.~\ref{fig:microvalidation}(b)]. Including it changes the
population-weighted benchmark by less than \(3\times10^{-4}~\mathrm{meV}\),
from 15.6234 to \(16.3106~\mathrm{meV}\) across the sweep. The calculation
therefore verifies the fixed-spatial-factor approximation at the precision
relevant to the reduced model; it does not provide a parameter-free absolute
coupling because the scale remains anchored at \(g_{\mathrm{fit}}(0)\).

\begin{table*}[t]
\caption{Parameter and unit convention used throughout the main text and
Supplemental Material. Angular-frequency quantities are used only in the
second-order Lorentz representation and time-domain TCMT; all pole, zero, and
fit equations are evaluated in energy units.}
\label{tab:conventions}
\centering
\scriptsize
\begin{tabular}{p{0.12\textwidth}
                p{0.39\textwidth}
                p{0.22\textwidth}
                p{0.18\textwidth}}
\toprule
Symbol & Physical meaning & Unit convention & Source or role\\
\midrule
\(F\) & Signed physical electric field along \(+\hat{\mathbf z}\) & \(\mathrm{kV\,cm^{-1}}\) & Independent sweep variable\\
\(E,\omega\) & Photon energy and angular frequency, related by \(E=\hbar\omega\) & meV; \(\mathrm{rad\,s^{-1}}\) & Spectral variables\\
\(E_{12}(F)\) & Intersubband transition energy \(E_2-E_1\) & meV & S--P output\\
\(E_{\mathrm{pc}}\) & Effective bare-photonic energy of the fitted TCMT mode & meV & Fixed after the zero-field fit\\
\(E_{\mathrm{bc}}\) & Intersubband-off RCWA reflection-minimum energy & meV & Post-fit full-wave check; not \(E_{\mathrm{pc}}\)\\
\(N_s\) & Donor sheet density per microscopic period & \(\mathrm{cm^{-2}}\) (reported) & Structure input\\
\(N_v(F)\) & Actual population-difference volume density, \([N_1(F)-N_2(F)]/L_p\) & \(\mathrm{m^{-3}}\) & RCWA susceptibility\\
\(f_{12}(F)\) & Dimensionless oscillator strength & dimensionless & S--P output\\
\(m_{12}^{\ast}(F)\) & Transition reference mass used in \(f_{12}\) & kg in equations; reported in \(m_0\) & Distinct from \(m^\ast(z,E)\)\\
\(W_{12}^{\mathrm{FWHM}}(F)\) & Matter intensity linewidth in energy units & meV & Dephasing-model output\\
\(\gamma_{12}(F)\) & Matter amplitude HWHM energy, \(W_{12}^{\mathrm{FWHM}}/2\) & meV & TCMT denominator\\
\(\Gamma_{12}^{\mathrm{L}}(F)\) & Lorentz angular-frequency damping, \(2\gamma_{12}/\hbar\) & \(\mathrm{s^{-1}}\) & Lorentz denominator only\\
\(\gamma_e\) & External-port amplitude HWHM energy & meV & Fixed zero-field fit\\
\(\gamma_{\mathrm{pc},\mathrm{nr}}\) & Nonradiative photonic amplitude HWHM energy & meV & Fixed zero-field fit\\
\(\gamma_{\mathrm{pc}}\) & Total photonic HWHM, \(\gamma_e+\gamma_{\mathrm{pc},\mathrm{nr}}\) & meV & Derived fixed quantity\\
\(g_{\mathrm{fit}}(F)\) & Effective fitted TCMT coupling energy & meV & Sole field-dependent fit parameter\\
\(g_{\mathrm{mic}}(F)\) & Population-, oscillator-strength-, mass-, and overlap-based benchmark & meV & Post-fit benchmark\\
\(\eta_S(F)\) & Zero-field-normalized residual spatial-overlap correction & dimensionless & Post-fit audit\\
\(\phi_c,\alpha\) & Common phase offset and phase slope & rad in exponent; degrees reported & Fixed zero-field fit\\
\(\tau_{\mathrm{ref}}\) & Common reference-plane delay, \(\hbar\alpha\) with \(\alpha\) in rad/energy & fs & Derived fixed quantity\\
\(S_{11},r,R\) & RCWA amplitude, TCMT amplitude, and reflectance \(|r|^2\) & dimensionless & Scattering observables\\
\(\mathrm{RMSE}_r,\mathrm{RMSE}_R\) & Complex-amplitude and reflectance root-mean-square errors & dimensionless & Fit diagnostics\\
\bottomrule
\end{tabular}
\end{table*}

\section{Field-resolved TCMT calibration and microscopic benchmark}

\begin{figure*}[t]
\centering
\includegraphics[width=0.94\textwidth]{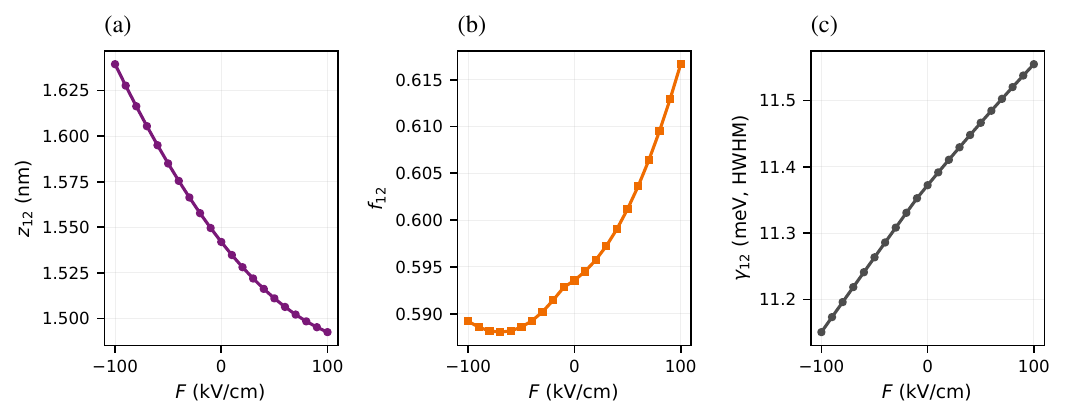}
\caption{Field-resolved quantum inputs used in the TCMT analysis.
(a) Dipole matrix element \(z_{12}\).
(b) Oscillator strength \(f_{12}\). (c) Optical-coherence energy HWHM
\(\gamma_{12}\) used directly in the TCMT denominator and converted to
\(\Gamma_{12}^{\mathrm{L}}=2\gamma_{12}/\hbar\) for the Lorentz susceptibility.}
\label{fig:qcseinputs}
\end{figure*}

Fig.~\ref{fig:qcseinputs} records the primitive quantum quantities entering
the microscopic coupling benchmark in main-text Fig.~2(b). The benchmark
also includes the field-resolved population difference and transition
reference mass. It is compared with the effective fitted couplings
\(g_{\mathrm{fit}}(F)\) listed in Table~\ref{tab:fit}. The linewidth shown
here is used consistently in both the RCWA tensor and TCMT denominator.

The \(F=0\) complex spectrum is fitted first. This baseline determines the
shared values \(E_{\mathrm{pc}}=196.548~\mathrm{meV}\),
\(\gamma_e=15.298~\mathrm{meV}\),
\(\gamma_{\mathrm{pc},\mathrm{nr}}=4.733~\mathrm{meV}\),
\(\phi_c=10.759^\circ\), and
\(\alpha=0.0712^\circ\,\mathrm{meV}^{-1}
=1.2423\times10^{-3}~\mathrm{rad\,meV^{-1}}\), together with
\(g_{\mathrm{fit}}(0)=15.910~\mathrm{meV}\). The phase slope corresponds to a common
reference delay of \(0.818~\mathrm{fs}\); \(\phi_c\) and \(\alpha\) are
converted to radians before evaluation of any complex exponential.

With \(E_{\mathrm{ref}}=200~\mathrm{meV}\), the common reference-plane phase
is \(\vartheta(E)=\phi_c+\alpha(E-E_{\mathrm{ref}})\). The baseline parameters
are obtained from the explicit complex-amplitude objective
\begin{equation}
\begin{aligned}
\bm{\theta}_0
&=\underset{\bm{\theta}}{\operatorname{arg\,min}}
\sum_{i=1}^{N_\lambda}
\left|S_{11}^{\mathrm{RCWA}}(\lambda_i,0)
-e^{i\vartheta(E_i)}r_{\mathrm{TCMT}}(\lambda_i,0)\right|^2\\
\bm{\theta}
&=(E_{\mathrm{pc}},g_{\mathrm{fit}}(0),\gamma_e,
\gamma_{\mathrm{pc},\mathrm{nr}},\phi_c,\alpha)
\end{aligned}
\label{eq:phase_objective_s}
\end{equation}
The multiplicative phase is nonzero everywhere and therefore cannot create
or remove a reflection zero. At nonzero field, only \(g_{\mathrm{fit}}(F)\)
is optimized; continuation proceeds from zero independently toward positive
and negative fields using the neighboring solution as the next initial value.

\begin{center}
\refstepcounter{table}\label{tab:shared_baseline}
\parbox{0.96\columnwidth}{\small\textbf{TABLE~\thetable.}\ Zero-field numerical values and the shared one-port TCMT baseline. The photonic and phase-delay parameters are fixed for all 21 field-resolved fits. The total photonic HWHM is \(\gamma_{\mathrm{pc}}=\gamma_e+\gamma_{\mathrm{pc},\mathrm{nr}}\). Phase quantities are reported in degrees but converted to radians in complex exponentials.}
\begin{tabular}{lrc}
\toprule
Quantity & Value & Unit\\
\midrule
\(E_{12}\) & 199.2820 & meV\\
\(\gamma_{12}\) & 11.3720 & meV\\
\(E_{\mathrm{bc}}\) & 196.4570 & meV\\
\(E_{\mathrm{pc}}\) & 196.5478 & meV\\
\(g_{\mathrm{fit}}(0)\) & 15.9097 & meV\\
\(\gamma_e\) & 15.2975 & meV\\
\(\gamma_{\mathrm{pc},\mathrm{nr}}\) & 4.7326 & meV\\
\(\gamma_{\mathrm{pc}}\) & 20.0301 & meV\\
\(\phi_c\) & 10.7587 & \({}^\circ\)\\
\(\alpha\) & 0.07118 & \({}^\circ\,\mathrm{meV}^{-1}\)\\
\(\tau_{\mathrm{ref}}\) & 0.8177 & fs\\
\(\mathrm{RMSE}_R\) & 0.0280 & dimensionless\\
\(\mathrm{RMSE}_r\) & 0.0260 & dimensionless\\
\({\cal C}\) & 1.111 & dimensionless\\
\bottomrule
\end{tabular}
\end{center}

The quantities in Table~\ref{tab:fit} fall into three distinct classes.
First, \(E_{\mathrm{pc}}\), \(\gamma_e\),
\(\gamma_{\mathrm{pc},\mathrm{nr}}\), \(\phi_c\), and \(\alpha\) are
baseline-calibrated parameters that are held fixed after the \(F=0\) fit.
The metasurface geometry and external port are unchanged during the
electrical sweep. The fixed values define one common effective photonic
baseline as an identifiability constraint. The two phase quantities define a
common de-embedding baseline between RCWA and TCMT. Together these fixed
quantities separate the shared photonic and phase-delay response from the
field-dependent quantum-well contribution.

Second, \(E_{12}(F)\) and \(\gamma_{12}(F)\) enter at each field from the
S--P calculation and effective dephasing model. They provide the
physics-informed link between the quantum calculation and the reduced model.

Third, \(g_{\mathrm{fit}}(F)\) is the only field-dependent fit parameter. Its
zero-field value is obtained in the initial joint fit, after which it is continued
sequentially from zero toward positive and negative bias. This single degree
of freedom accommodates the effective field dependence not represented by
the fixed microscopic-strength benchmark. It therefore serves as the
hybridization parameter of the reduced representation without being
identified with an independently measured near-field overlap. Sequential
continuation uses the neighboring-field solution as the next initial value,
suppressing discontinuous jumps among local fit minima.
Table~\ref{tab:fit} gives all 21 fits.

\begin{table*}[t]
\caption{Field-resolved one-port TCMT fits over
\(5.0\)--\(7.6~\mu\mathrm{m}\). Energies and HWHM rates are in meV,
\(F\) is in \(\mathrm{kV\,cm^{-1}}\), and RMSE values are dimensionless.
The shared baseline parameters in Table~\ref{tab:shared_baseline} are held
fixed; \(E_{12}\) and \(\gamma_{12}\) are supplied by the S--P calculation
and optical-dephasing model, respectively, and
\(g_{\mathrm{fit}}\) is the only field-dependent effective fit parameter.
The RMSE definitions are given in Sec.~\ref{sec:rmse}.}
\label{tab:fit}
\centering
\begin{tabular}{rrrrrr}
\toprule
\(F\) & \(E_{12}\) & \(\gamma_{12}\) & \(g_{\mathrm{fit}}\) &
\(\mathrm{RMSE}_R\) & \(\mathrm{RMSE}_r\)\\
\midrule
-100 & 173.821 & 11.151 & 14.470 & 0.0320 & 0.0268 \\
-90 & 176.308 & 11.174 & 14.590 & 0.0315 & 0.0267 \\
-80 & 178.822 & 11.196 & 14.715 & 0.0309 & 0.0264 \\
-70 & 181.361 & 11.219 & 14.847 & 0.0303 & 0.0262 \\
-60 & 183.923 & 11.241 & 14.985 & 0.0298 & 0.0259 \\
-50 & 186.508 & 11.264 & 15.131 & 0.0293 & 0.0257 \\
-40 & 189.114 & 11.286 & 15.284 & 0.0289 & 0.0255 \\
-30 & 191.739 & 11.308 & 15.443 & 0.0285 & 0.0255 \\
-20 & 194.383 & 11.331 & 15.608 & 0.0283 & 0.0256 \\
-10 & 196.985 & 11.353 & 15.771 & 0.0281 & 0.0258 \\
0 & 199.282 & 11.372 & 15.910 & 0.0280 & 0.0260 \\
+10 & 201.557 & 11.392 & 16.048 & 0.0280 & 0.0263 \\
+20 & 203.814 & 11.411 & 16.185 & 0.0281 & 0.0266 \\
+30 & 206.063 & 11.429 & 16.323 & 0.0282 & 0.0269 \\
+40 & 208.307 & 11.448 & 16.460 & 0.0284 & 0.0273 \\
+50 & 210.542 & 11.466 & 16.596 & 0.0286 & 0.0276 \\
+60 & 212.768 & 11.485 & 16.732 & 0.0288 & 0.0279 \\
+70 & 214.981 & 11.502 & 16.867 & 0.0290 & 0.0282 \\
+80 & 217.179 & 11.520 & 17.000 & 0.0293 & 0.0284 \\
+90 & 219.361 & 11.538 & 17.133 & 0.0295 & 0.0286 \\
+100 & 221.523 & 11.555 & 17.264 & 0.0298 & 0.0288 \\
\bottomrule
\end{tabular}
\end{table*}

The reflectance RMSE ranges from 0.0280 to 0.0320, and the complex-amplitude
RMSE from 0.0255 to 0.0288. A single \(\phi_c\), \(\alpha\), and one shared
set of photonic parameters are used for every field.

The calibrated field dependence is then tested against the microscopic
susceptibility strength. To determine how much follows directly from the
microscopic susceptibility numerator, we define
\(\mathcal{S}(F)=[N_1(F)-N_2(F)]f_{12}(F)/m_{12}^{\ast}(F)\) and evaluate
\begin{equation}
g_{\mathrm{mic}}(F)=g_{\mathrm{fit}}(0)
\left[\frac{\mathcal{S}(F)}{\mathcal{S}(0)}\right]^{1/2}
\eta_S(F)
\label{eq:g_f_benchmark}
\end{equation}
The scale is fixed at \(F=0\), and no parameter is fitted away from that
field. The post-fit audit in Sec.~\ref{sec:microaudit} shows that the overlap
factor changes this trajectory by less than
\(3\times10^{-4}~\mathrm{meV}\). Over all 21 fields the field-resolved fit gives aggregate
complex-amplitude and reflectance RMSE values of 0.02680 and 0.02923,
respectively. The microscopic benchmark gives 0.03875 and 0.02915. The
intensity error is therefore essentially unchanged, while the complex error
increases by 45\%. The phase-resolved far field, rather than reflectance
alone, distinguishes the additional effective field dependence carried by
\(g_{\mathrm{fit}}(F)\). Because the explicitly evaluated spatial correction
is negligible, it does not explain the remaining difference; that difference
quantifies what the microscopic strength-and-overlap benchmark does not
represent in the reduced complex response.

\section{Identifiability and fitting-window controls}

\begin{table*}[t]
\caption{Zero-field complex-response reduction and reflectance-only control.
The reflectance-only result is aligned afterward by its best common constant
and linear phase solely to evaluate \(\mathrm{RMSE}_r\). Energies and rates
are in meV.}
\label{tab:identifiability}
\centering
\begin{tabular}{lcccccc}
\toprule
Fit observable & \(E_{\mathrm{pc}}\) & \(g_{\mathrm{fit}}(0)\) & \(\gamma_e\) &
\(\gamma_{\mathrm{pc},\mathrm{nr}}\) & \(\mathrm{RMSE}_R\) &
\(\mathrm{RMSE}_r\)\\
\midrule
Complex \(S_{11}\) & 196.548 & 15.910 & 15.298 & 4.733 & 0.0280 & 0.0260\\
Reflectance only & 197.897 & 15.166 & 14.624 & 5.939 & 0.0181 & 0.0562\\
\bottomrule
\end{tabular}
\end{table*}

The zero-field complex fit was repeated from 100 randomized initial points.
The starting ranges were 180--215 meV for \(E_{\mathrm{pc}}\), 5--30 meV for
\(g_{\mathrm{fit}}(0)\), 3--30 meV for \(\gamma_e\), 0.2--15 meV for
\(\gamma_{\mathrm{pc},\mathrm{nr}}\), \(-30^\circ\)--\(30^\circ\) for
\(\phi_c\), and \(-0.5\)--\(0.5^\circ\,\mathrm{meV}^{-1}\) for
\(\alpha\). Every run converged to the same solution:
\(E_{\mathrm{pc}}=196.54784\), \(g_{\mathrm{fit}}(0)=15.90973\),
\(\gamma_e=15.29753\), and
\(\gamma_{\mathrm{pc},\mathrm{nr}}=4.73258~\mathrm{meV}\), with
\(\mathrm{RMSE}_r=0.0259812\). The maximum spread of any reported energy or
rate was below \(5\times10^{-8}~\mathrm{meV}\). This multi-start test
establishes practical uniqueness within the stated fitting domain.

\begin{table*}[t]
\caption{Zero-field complex-fit sensitivity to the wavelength window.
Energies and rates are in meV, \(\phi_c\) is in degrees, \(\alpha\) is in
degrees per meV, \(N_\lambda\) is a sample count, and RMSE values are
dimensionless.}
\label{tab:window}
\centering
\begin{tabular}{ccccccccc}
\toprule
Window (\(\mu\mathrm{m}\)) & \(N_\lambda\) & \(E_{\mathrm{pc}}\) & \(g_{\mathrm{fit}}(0)\) &
\(\gamma_e\) & \(\gamma_{\mathrm{pc},\mathrm{nr}}\) & \(\phi_c\) &
\(\alpha\) & \(\mathrm{RMSE}_r\)\\
\midrule
5.0--7.6 & 217 & 196.548 & 15.910 & 15.298 & 4.733 & 10.759 & 0.0712 & 0.02598\\
5.2--7.4 & 184 & 196.547 & 15.889 & 15.368 & 4.812 & 11.435 & 0.0668 & 0.02438\\
5.4--7.2 & 150 & 196.480 & 15.877 & 15.450 & 4.873 & 12.423 & 0.0614 & 0.02087\\
5.5--7.1 & 134 & 196.410 & 15.867 & 15.479 & 4.896 & 13.025 & 0.0530 & 0.01866\\
\bottomrule
\end{tabular}
\end{table*}

The extracted coupling changes by at most \(0.043~\mathrm{meV}\) over these
windows. The systematic variation of the phase baseline and external rate
accounts for the smooth far-off-resonant background, while the hybridization
parameter remains stable.

\section{Complex reflection zeros, critical coupling, and linewidth sensitivity}

The internal poles satisfy
\begin{equation}
\left[E-E_{\mathrm{pc}}+i\gamma_{\mathrm{pc}}\right]
\left[E-E_{12}(F)+i\gamma_{12}(F)\right]-g_{\mathrm{fit}}^2(F)=0
\label{eq:s_pole_equation}
\end{equation}
In the lossless limit, their real energies reduce to
\begin{align}
\Delta(F)&=E_{12}(F)-E_{\mathrm{pc}}\nonumber\\
E_\pm(F)&=\frac{E_{\mathrm{pc}}+E_{12}(F)}{2}\nonumber\\
&\quad\pm\sqrt{g_{\mathrm{fit}}^2(F)+\frac{\Delta^2(F)}{4}}
\label{eq:s_lossless_branches}
\end{align}
At exact resonance, nonzero real pole splitting requires
\begin{equation}
g_{\mathrm{fit}}(F)>
\frac{|\gamma_{\mathrm{pc}}-\gamma_{12}(F)|}{2}
\label{eq:s_pole_split}
\end{equation}
whereas the more conservative resolvability benchmark and cooperativity are
\begin{equation}
g_{\mathrm{fit}}(F)>
\frac{\gamma_{\mathrm{pc}}+\gamma_{12}(F)}{2}
\qquad
{\cal C}(F)=
\frac{g_{\mathrm{fit}}^2(F)}
{\gamma_{\mathrm{pc}}\gamma_{12}(F)}
\label{eq:s_strong_criteria}
\end{equation}
These internal-mode criteria do not determine the external one-port coupling
regime.

The one-port TCMT coefficient in the main text is
\[
r(E,F)=-1+\frac{2\gamma_e}{D(E,F)}
\]
Its two complex zeros are obtained from
\begin{equation}
\begin{split}
&\left[E-E_{\mathrm{pc}}
+i(\gamma_{\mathrm{pc},\mathrm{nr}}-\gamma_e)\right]\\
&\qquad\times
\left[E-E_{12}(F)+i\gamma_{12}(F)\right]-g_{\mathrm{fit}}^2(F)=0
\end{split}
\label{eq:s_zero}
\end{equation}
The two roots of Eq.~(\ref{eq:s_zero}) are the LP- and UP-associated
reflection zeros of the same hybridized two-mode system, rather than two
independently coupled resonances. Their separate trajectories therefore
provide a branch-resolved analysis of one cavity--ISBT coupling problem.
All quantities in Eq.~(\ref{eq:s_zero}) are taken from the constrained
field-resolved calibration: \(E_{\mathrm{pc}}\), \(\gamma_e\), and
\(\gamma_{\mathrm{pc},\mathrm{nr}}\) remain fixed; \(E_{12}(F)\) and
\(\gamma_{12}(F)\) are fixed by the S--P calculation and optical model; and
\(g_{\mathrm{fit}}(F)\) is the sole field-dependent fit parameter. The zero
calculation uses this calibrated parameter set.

At simultaneous cavity--matter resonance,
\(E=E_{\mathrm{pc}}=E_{12}\), the general complex-zero condition reduces to
\begin{equation}
\gamma_e=\gamma_{\mathrm{pc},\mathrm{nr}}+
\frac{g_{\mathrm{fit}}^2(F)}{\gamma_{12}(F)}
\label{eq:s_critical_resonance}
\end{equation}
This is a special resonant balance, not a field-independent criterion away
from resonance. In the general detuned case, critical coupling is located by
the real-axis crossing of the complete complex roots of
Eq.~(\ref{eq:s_zero}).

For the main-text \(\exp(-i\omega t)\) convention, zeros with negative and
positive imaginary energies represent the undercoupled and overcoupled
one-port reflection regimes, respectively. A zero with zero imaginary energy
satisfies the critical-coupling condition separating those regimes. This
classification is distinct from strong versus weak coherent
light--matter coupling, which is assessed from the poles, linewidth ratio,
and cooperativity.

\begin{table*}[t]
\caption{Complete field-resolved complex reflection zeros calculated from
the constrained TCMT parameters in Tables~\ref{tab:shared_baseline} and
\ref{tab:fit}. Energies are in meV and \(F\) is in
\(\mathrm{kV\,cm^{-1}}\). ``Under'' and ``over'' denote the undercoupled and
overcoupled one-port reflection regimes under the
\(\exp(-i\omega t)\) convention. The LP and UP columns together report all
42 complex zeros used in main-text Fig.~4.}
\label{tab:complex_zeros}
\centering
\begin{tabular}{rrrrrrr}
\toprule
& \multicolumn{3}{c}{Lower polariton} &
\multicolumn{3}{c}{Upper polariton}\\
\cmidrule(lr){2-4}\cmidrule(lr){5-7}
\(F\) & \(\operatorname{Re}E_{z,\mathrm{LP}}\) &
\(\operatorname{Im}E_{z,\mathrm{LP}}\) & Regime &
\(\operatorname{Re}E_{z,\mathrm{UP}}\) &
\(\operatorname{Im}E_{z,\mathrm{UP}}\) & Regime\\
\midrule
-100 & 168.576 & -7.722 & under & 201.792 & 7.136 & over\\
-90 & 170.735 & -7.313 & under & 202.121 & 6.705 & over\\
-80 & 172.881 & -6.830 & under & 202.488 & 6.199 & over\\
-70 & 175.004 & -6.256 & under & 202.905 & 5.602 & over\\
-60 & 177.085 & -5.572 & under & 203.386 & 4.896 & over\\
-50 & 179.099 & -4.758 & under & 203.956 & 4.059 & over\\
-40 & 181.005 & -3.795 & under & 204.656 & 3.074 & over\\
-30 & 182.743 & -2.678 & under & 205.544 & 1.935 & over\\
-20 & 184.239 & -1.438 & under & 206.692 & 0.673 & over\\
-10 & 185.420 & -0.183 & under & 208.112 & -0.605 & under\\
0 & 186.239 & 0.881 & over & 209.591 & -1.688 & under\\
+10 & 186.872 & 1.844 & over & 211.234 & -2.671 & under\\
+20 & 187.369 & 2.693 & over & 212.993 & -3.539 & under\\
+30 & 187.775 & 3.435 & over & 214.836 & -4.299 & under\\
+40 & 188.117 & 4.080 & over & 216.738 & -4.964 & under\\
+50 & 188.412 & 4.643 & over & 218.678 & -5.544 & under\\
+60 & 188.674 & 5.134 & over & 220.641 & -6.054 & under\\
+70 & 188.909 & 5.565 & over & 222.619 & -6.502 & under\\
+80 & 189.123 & 5.944 & over & 224.604 & -6.899 & under\\
+90 & 189.319 & 6.278 & over & 226.589 & -7.251 & under\\
+100 & 189.500 & 6.575 & over & 228.571 & -7.565 & under\\
\bottomrule
\end{tabular}
\end{table*}
Table~\ref{tab:complex_zeros} gives the complete field-resolved zero
trajectories. The lower-polariton zero changes from
undercoupled to overcoupled between \(-10\) and
\(0~\mathrm{kV\,cm^{-1}}\), whereas the upper-polariton zero changes from
overcoupled to undercoupled between \(-20\) and
\(-10~\mathrm{kV\,cm^{-1}}\). Linear interpolation of the TCMT zero
imaginary parts gives the following critical fields.
\[
\begin{aligned}
F_{\mathrm{c,LP}}^{\mathrm{TCMT}}
&=-8.279~\mathrm{kV\,cm^{-1}}\\
F_{\mathrm{c,UP}}^{\mathrm{TCMT}}
&=-14.734~\mathrm{kV\,cm^{-1}}
\end{aligned}
\]
The associated interpolated real energies are 185.561 and
\(207.440~\mathrm{meV}\), respectively. These decimal values are
model-conditioned interpolation estimates; for the nominal TCMT--RCWA
comparison, the more robust results are the sampled-field brackets.
The raw RCWA spectra provide an independent complex check. Within every
adjacent field--wavelength cell, we bilinearly interpolate \(S_{11}\) and
solve both zero conditions simultaneously.
\[
\operatorname{Re}S_{11}=0,\qquad \operatorname{Im}S_{11}=0
\]
The solutions are
\[
\begin{aligned}
F_{\mathrm{LP}}&=-9.412~\mathrm{kV\,cm^{-1}}\\
\lambda_{\mathrm{LP}}&=6.7031~\mu\mathrm{m}&
E_{\mathrm{LP}}&=184.966~\mathrm{meV}
\end{aligned}
\]
and
\[
\begin{aligned}
F_{\mathrm{UP}}&=-16.002~\mathrm{kV\,cm^{-1}}\\
\lambda_{\mathrm{UP}}&=6.0005~\mu\mathrm{m}&
E_{\mathrm{UP}}&=206.624~\mathrm{meV}
\end{aligned}
\]
The interpolated solutions satisfy both components of the complex-zero
condition within numerical precision. The sampled real-frequency full-wave
coefficients and reduced description place zeros in the same two
sampled-field intervals. The RCWA-minus-TCMT differences are
\(-1.13~\mathrm{kV\,cm^{-1}}\) for the lower-polariton feature and
\(-1.27~\mathrm{kV\,cm^{-1}}\) for the upper-polariton feature.
Accordingly, the sub-grid values are reported as interpolation estimates
within the sampled-field intervals.

At the directly sampled field \(F=-10~\mathrm{kV\,cm^{-1}}\), the lower-
and upper-polariton sampled-grid RCWA minima are \(5.6601\times10^{-5}\) and
\(9.7008\times10^{-4}\). This sampled spectrum exhibits near-critical coupling, and
interpolation of the complex coefficients locates the real-axis crossings. The
robustness of these zero crossings to the assumed matter linewidth is assessed
by repeating the calibration after scaling the complete
\(\gamma_{12}(F)\) trajectory.

\FloatBarrier

\begin{samepage}
\begin{center}
\refstepcounter{table}\label{tab:linewidth_sensitivity}
\parbox{0.96\columnwidth}{\small\textbf{TABLE~\thetable.}
Sensitivity to a common scale factor applied to \(\gamma_{12}(F)\).
The scale factor \(s_\gamma\) is dimensionless. Rates and energies are in meV;
RMSE values are dimensionless and critical fields are in
\(\mathrm{kV\,cm^{-1}}\). The critical fields are linear
interpolations of the refitted TCMT zero trajectories.}
\small
\begin{tabular}{lrrr}
\toprule
Quantity & \(s_\gamma=0.8\) & \(s_\gamma=1.0\) & \(s_\gamma=1.2\)\\
\midrule
\(E_{\mathrm{pc}}\) & 196.453 & 196.548 & 196.627\\
\(g_{\mathrm{fit}}(0)\) & 14.555 & 15.910 & 17.261\\
\(\gamma_e\) & 15.732 & 15.298 & 14.907\\
\(\gamma_{\mathrm{pc},\mathrm{nr}}\) & 6.454 & 4.733 & 3.119\\
\(\mathrm{RMSE}_r\) & 0.0390 & 0.0268 & 0.0384\\
\(\min_F[2g_{\mathrm{fit}}(F)/(\gamma_{\mathrm{pc}}+\gamma_{12}(F))]\) & 0.869 & 0.928 & 0.987\\
\(\min_F{\cal C}(F)\) & 0.923 & 0.937 & 0.996\\
\(F_{\mathrm{c,LP}}\) & -12.96 & -8.28 & -4.08\\
\(F_{\mathrm{c,UP}}\) & -11.11 & -14.73 & -17.63\\
\bottomrule
\end{tabular}
\end{center}
\end{samepage}

Table~\ref{tab:linewidth_sensitivity} reports the sensitivity analysis for
scales \(s_\gamma=0.8,1.0,\) and \(1.2\) applied to every nominal
\(\gamma_{12}(F)\).
For each scale, the \(F=0\) complex spectrum is refitted for
\(E_{\mathrm{pc}}\), \(g_{\mathrm{fit}}(0)\), \(\gamma_e\),
\(\gamma_{\mathrm{pc},\mathrm{nr}}\), \(\phi_c\), and \(\alpha\). Those
shared parameters are then held fixed while only \(g_{\mathrm{fit}}(F)\) is continued over
the positive and negative field branches. Each scale therefore follows the
full nominal calibration protocol.

The two separated RCWA minima and their avoided-crossing topology are
unchanged because the reference full-wave spectra are unchanged. The
refitted reduced model also retains comparable complex-response accuracy,
although the nominal linewidth gives the lowest aggregate
\(\mathrm{RMSE}_r\). In contrast, the conservative threshold and the
interpolated TCMT critical fields shift appreciably. The linewidth study
therefore establishes that the qualitative anticrossing and near-threshold
hybridization persist across the tested range, while the binary threshold
classification and interpolated TCMT critical fields shift with the
linewidth calibration.

\section{TCMT rate convention and complex-residual definition}
\label{sec:rmse}

For monochromatic excitation, the main-text time-domain equations give, in
energy notation,
\begin{align}
\left[\gamma_{\mathrm{pc}}-i(E-E_{\mathrm{pc}})\right]a
-ig_{\mathrm{fit}}(F)b
&=\sqrt{2\hbar\gamma_e}\,s_+
\label{eq:s_ssa}\\
\left[\gamma_{12}(F)-i(E-E_{12}(F))\right]b
-ig_{\mathrm{fit}}(F)a&=0
\label{eq:s_ssb}
\end{align}
The corresponding energy-domain input--output relation is
\(s_-=-s_++\sqrt{2\gamma_e/\hbar}\,a\). The factors of \(\sqrt{\hbar}\)
cancel in the dimensionless ratio \(s_-/s_+\). Eliminating \(b\) gives
\begin{align}
b&=\frac{ig_{\mathrm{fit}}(F)}
{\gamma_{12}(F)-i[E-E_{12}(F)]}\,a
\label{eq:s_b_of_a}\\
a&=\frac{\sqrt{2\hbar\gamma_e}\,s_+}{D(E,F)}
\label{eq:s_a_solution}\\
D(E,F)&=\gamma_{\mathrm{pc}}-i(E-E_{\mathrm{pc}})
+\frac{g_{\mathrm{fit}}^2(F)}
{\gamma_{12}(F)-i[E-E_{12}(F)]}
\label{eq:s_D}
\end{align}
Substitution into the input--output relation yields the main-text
\(r(E,F)=-1+2\gamma_e/D(E,F)\). The reflection phase is
\(\operatorname{unwrap}\{\arg r\}\); it is undefined at an exact zero and is
interpreted through the limiting phase on either side.

The main text uses amplitude HWHM energies. If a fitted or measured intensity
linewidth is reported as the energy full width at half maximum
\(W_j^{\mathrm{FWHM}}\), the value used in the TCMT denominator is
\begin{equation}
\gamma_j=\frac{W_j^{\mathrm{FWHM}}}{2}
\end{equation}
Likewise, the angular-frequency amplitude-decay rate is
\(\kappa_j=\gamma_j/\hbar\) and
\(g_\omega(F)=g_{\mathrm{fit}}(F)/\hbar\). These conversions must
be applied consistently
to the cavity and matter modes before evaluating any strong-coupling
inequality.

For the global field-resolved comparison, the complex residual after the
common phase-delay de-embedding is defined with the same sign as main-text
Fig.~3(c):
\begin{equation}
\varepsilon_r(E,F)=
e^{i[\phi_c+\alpha(E-E_{\mathrm{ref}})]}r_{\mathrm{TCMT}}(E,F)
-S_{11}^{\mathrm{RCWA}}(E,F)
\end{equation}
The same \(\phi_c\) and \(\alpha\) are used for the entire map. For the \(N_\lambda\)
samples in the common \(5.0\)--\(7.6~\mu\mathrm{m}\) fitting window, let
\(R_i^{\mathrm{RCWA}}=|S_{11}^{\mathrm{RCWA}}(\lambda_i,F)|^2\) and
\(R_i^{\mathrm{TCMT}}=|r_{\mathrm{TCMT}}(\lambda_i,F)|^2\). The
dimensionless fit-quality measures in Table~\ref{tab:fit} are
\begin{align}
\mathrm{RMSE}_R(F)
&=\left[
\frac{1}{N_\lambda}\sum_{i=1}^{N_\lambda}
\left(R_i^{\mathrm{TCMT}}-R_i^{\mathrm{RCWA}}\right)^2
\right]^{1/2}
\label{eq:rmse_reflectance}\\
\mathrm{RMSE}_r(F)
&=\left[
\frac{1}{N_\lambda}\sum_{i=1}^{N_\lambda}
\left|\varepsilon_r(\lambda_i,F)\right|^2
\right]^{1/2}
\label{eq:rmse_complex}
\end{align}
Thus, \(\mathrm{RMSE}_R\) measures the reflectance mismatch, whereas
\(\mathrm{RMSE}_r\) is sensitive to both amplitude and phase.

\end{document}